# Magneto-elastic coupling and magnetocaloric effect in hexagonal Mn-Fe-P-Si compounds

Nguyen Huu Dung,[1]* Lian Zhang,[1,2] Zhi Qiang Ou,[1] Ekkes Brück[1]

[1] *Fundamental Aspects of Materials and Energy (FAME), Faculty of Applied Sciences, Delft University of Technology, 2629 JB Delft, Netherlands*

[2] *BASF Netherlands B.V., 3454 PK De Meern, Netherlands*

*Email: h.d.nguyen@tudelft.nl

Structural, magnetic and magnetocaloric properties of $Fe_2P$-based Mn-Fe-P-Si compounds were investigated. The study reveals a large magneto-elastic coupling that starts to develop in the paramagnetic state and grows when the ferromagnetic transition temperature is approached. Based on the behavior of the magneto-elastic coupling, we show the thermal evolution of the magnetic moments. On cooling, magnetic moments on the tetrahedral site form and gradually increase in the paramagnetic state. At the magnetic ordering temperature the moments attain a much larger value in a discontinuous step. We also find that the hysteresis and magnetic entropy change are correlated with discontinuous changes in the lattice parameters at the transition temperature. Small hysteresis can be obtained while maintaining giant magnetocaloric effect.

## I. INTRODUCTION

Magnetocaloric materials are the key to eco-friendly magnetic refrigeration technology.[1-3] Advanced magnetocaloric materials take advantage of a first-order magnetic transition (FOMT) because the FOMT is often associated with a change in crystal lattice which enhances magnetocaloric effects. Therefore the magnetocaloric materials often have magneto-elastic or magneto-structural transition.[4] However, thermal or magnetic hysteresis, which is intrinsic to a FOMT, is detrimental to the refrigeration-cycle efficiency.[5] Many efforts have been made to reduce the hysteresis of the magnetocaloric materials.[1,3-7]

Magneto-elastic coupling in a solid may result in a change of shape or volume due to an applied magnetic field or a change in temperature. The effect is often the most pronounced if the state of the solid is close to a phase transition. Usually the effect is small for most magnetic materials. However certain materials possess a large magneto-elastic coupling. These include some ferromagnetic shape memory,[8] multiferroic[9] and magnetocaloric materials.[10,11]

$Fe_2P$-based compounds display giant magnetocaloric effects (MCEs).[1,2,6,7] It is well-known that they often undergo a first-order magneto-elastic transition at which there are sudden changes in the $a$ and $c$ parameters ($\Delta a/a$ and $\Delta c/c$, respectively) but the volume hardly changes at the transition temperature.[1] Recently, $(Mn,Fe)_{1.95}(P,Si)$ compounds with hexagonal $Fe_2P$-type structure have been reported as promising materials for near room-temperature refrigeration applications because they display a giant MCE with very small hysteresis in a large temperature range.[7] Additionally, these compounds consist of abundant, cheap and non-toxic elements.

It is well-known for the hexagonal MnFe(P,Si) compounds that Mn preferentially occupies the $3g$ site while Fe favors the $3f$ site.[12,13] Although *ab initio* electronic-structure calculations predict that Si prefers the $2c$ site, while P tends to occupy the $1b$ site,[14] neutron diffraction experiments do not show a clear evidence for this.[12] Therefore, Mn($3g$) is surrounded by five P/Si nearest neighbors while Fe($3f$) has only four P/Si nearest neighbors forming a tetrahedron. Mn($3g$) and Fe($3f$) creates two basal planes in turn, alternating along the $c$ axis (see Fig. 1). Since the Mn($3g$) - P/Si distance is larger than the Fe($3f$) - P/Si distance and the Mn $3d$ electrons hybridize less than the Fe $3d$ electrons, the chemical bonding between the Fe($3f$) and P/Si is much stronger than that between Mn($3g$) and P/Si. The Mn($3g$) is, therefore, surrounded loosely by P/Si and its $3d$ electrons are more localized on the $3g$ site while the electron density is spatially extended over Fe($3f$) and P/Si. The moment of Mn($3g$) was found to be larger than that of Fe($3f$).[12,13]

Recently, mixed strong and weak magnetism in adjacent Mn($3g$) and Fe($3f$) lattice planes was proposed from *ab initio*

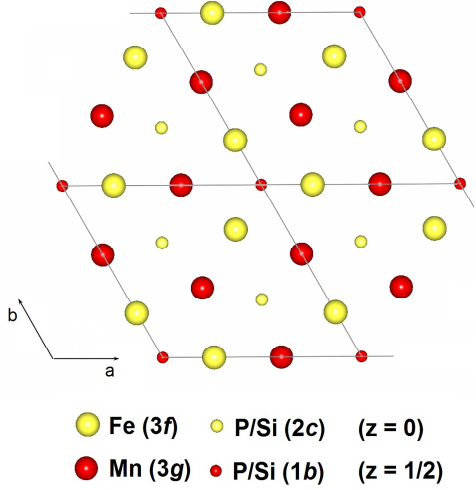

FIG. 1 (Color online). Arrangement of unit cells in hexagonal MnFe(P,Si) compounds.

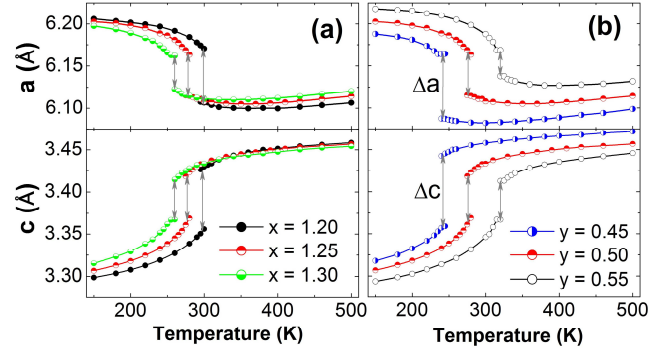

FIG. 2 (Color online). Lattice parameters $a$ and $c$ as a function of temperature derived from X-ray diffraction patterns measured in zero magnetic field upon heating for the $Mn_xFe_{1.95-x}P_{0.50}Si_{0.50}$ (a) and $Mn_{1.25}Fe_{0.70}P_{1-y}Si_y$ (b) compounds. The symbol size is larger than the error bar.

electronic structure calculations for the MnFe(P,Si) compounds.[7] The Fe(3$f$) 3$d$ states strongly hybridize with the nearest neighbors at the critical temperature, leading to the disappearance of Fe moments on the 3$f$ site, and simultaneously loss of ferromagnetic order. Since the Mn(3$g$) 3$d$ electrons are more localized, the magnetic transition only causes a small reduction of the Mn moments on the 3$g$ site. The FOMT therefore originates from the competition between the magnetic moments and chemical bonding at the 3$f$ sites. The Fe(3$f$) magnetic moments are metastable and sensitive to temperature, magnetic field and pressure. In the Mn-rich compounds studied here, excess Mn will enter this 3$f$ site replacing the metastable Fe moments.

Here we report on $Mn_xFe_{1.95-x}P_{1-y}Si_y$ ($x$ = 1.20 – 1.30, $y$ = 0.45 – 0.55) compounds with emphasis on the correlation between the magnetic and structural properties. The thermal evolution of the magneto-elastic coupling appears to be very sensitive to composition. Mixed magnetism and the detailed evolution of magnetic moments are proposed as the driving mechanism for the first-order magneto-elastic transition.

## II. EXPERIMENTAL DETAILS

$Mn_xFe_{1.95-x}P_{0.50}Si_{0.50}$ ($x$ = 1.20 - 1.30) and $Mn_{1.25}Fe_{0.70}P_{1-y}Si_y$ ($y$ = 0.45 - 0.55) compounds were prepared as described previously.[4,7] A SQUID magnetometer (Quantum Design MPMS 5XL) with the RSO mode was employed for the magnetic measurements. The X-ray diffraction patterns were collected at various temperatures in zero-field using a PANalytical X-pert Pro diffractometer equipped with an Anton Paar TTK450 low temperature camera using Cu K$\alpha$ radiation, a secondary beam flat crystal monochromator and a multichannel X'celerator detector. Starting at 150 K each X-ray pattern was recorded at a constant temperature and the following one was recorded at a higher temperature. Structure determination and refinement using FullProf[15] show that all the samples crystallize in the hexagonal $Fe_2P$-type structure (space group P-*62m*).

## III. MAGNETO-ELASTIC COUPLING AND MAGNETIC MOMENTS

From the X-ray patterns we derived the temperature dependence of the lattice parameters of the hexagonal unit cell as depicted in Fig. 2. The most prominent feature in Fig. 2 is the discontinuous jump in lattice parameters that shifts to lower and higher temperature with increasing $x$ and $y$, respectively. At these transitions we observe coexistence of two phases with the same hexagonal symmetry but different lattice parameters. All the samples show the same trend in the thermal evolution of the lattice parameters. Below the transition temperature ($T_C$),[16] the lattice parameters change in opposite sense, i.e., the $a$ parameter decreases while the $c$ parameter increases with increasing temperature. This happens gradually near and below $T_C$ and abruptly at the transition. The magnitude of the $\Delta a$ and $\Delta c$ decreases with both increasing Mn and Si content. While away from $T_C$, the $a$ and $c$ parameters change more gradually with respect to temperature for the samples with less Mn and Si. Figure 3 shows the temperature dependence of the magnetization measured in a magnetic field of 1 T for the $Mn_xFe_{1.95-x}P_{0.50}Si_{0.50}$ and $Mn_{1.25}Fe_{0.70}P_{1-y}Si_y$ compounds. Sharp ferro-paramagnetic transitions are observed for these samples.

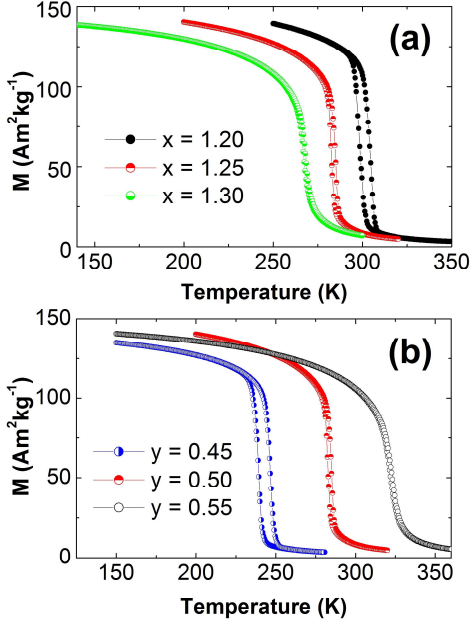

FIG. 3 (Color online). Temperature dependence of the magnetization measured in a magnetic field of 1 T upon cooling and heating for the $Mn_xFe_{1.95-x}P_{0.50}Si_{0.50}$ (a) and $Mn_{1.25}Fe_{0.70}P_{1-y}Si_y$ (b) compounds.

The thermal hysteresis can be attributed to the energy barrier associated with the FOMT. However, the thermal hysteresis can be reduced to negligible values as demonstrated for $x = 1.30$ in Fig. 3a and $y = 0.55$ in Fig. 3b. Note that these samples still show discontinuous changes in lattice parameters in Fig. 2.

In the picture of chemical bonding competing with magnetic moment put forward in Ref. 7 the behavior of the $a$ and $c$ parameters indicates that the loss of moment becomes more gradual starting already below $T_C$ with increasing temperature. As a result, the magnetic moments decrease upon heating, i.e., the Fe/Mn moments on the $3f$ site reduce markedly, while the Mn moments on the $3g$ site do not significantly change near $T_C$. This places similar behavior in hexagonal $Fe_2P$-based $(Mn,Fe)_2(P,Ge)$ compounds in a new light.[6] By neutron diffraction, Liu et al.[17] found a decrease in the Fe/Mn($3f$) moments upon heating near $T_C$ in $Mn_{1.1}Fe_{0.9}P_{0.8}Ge_{0.2}$ compound, while the Mn($3g$) moments do not significantly change. Additionally, they showed that the intralayer Fe/Mn($3f$) - P/Ge($2c$) distance drops strongly while the interlayer Fe/Mn($3f$) - P/Ge($1b$) distance expands a bit at the ferro-paramagnetic transition. Moreover, the intralayer Fe/Mn($3f$) - Fe/Mn($3f$) distance also decreases strongly, while the interlayer Fe/Mn($3f$) - Mn($3g$) distance hardly changes. A decrease in atomic distance leads to a reduction in the electron localization and magnetic moments. Thus, only the intralayer hybridization between the Fe/Mn($3f$) $3d$ states and the nearest neighbors contributes to the loss of the Fe/Mn($3f$) moments, the interlayer hybridization does not. These observations are true in a similar way for the Mn($3g$) moments. However, the displacement of the Mn($3g$) is much smaller than that of the Fe/Mn($3f$). Furthermore, the Mn($3g$) - P/Ge and intralayer Mn($3g$) - Mn($3g$) distances are larger than the Fe/Mn($3f$) - P/Ge and intralayer Fe/Mn($3f$) - Fe/Mn($3f$) distances, respectively, and the Mn $3d$ electrons are more localized. Thus, the loss of the Mn($3g$) moments is very small compared with that of the Fe/Mn($3f$) moments. In general, the decrease in the $a$ parameter reduces the intralayer atomic distances and the magnetic moments, and simultaneously enhances the intralayer bonding, especially for the Fe/Mn(3*f*) layers. This may be a common feature for all other hexagonal Fe$_2$P-based materials exhibiting a first-order magneto-elastic transition.

TABLE I. Magnetic-ordering temperature ($T_C$) in 1 T derived from the magnetization curves measured on heating, onset temperature of the moment formation on the $3f$ site ($T_O$), thermal hysteresis ($\Delta T_{hys}$) derived from the magnetization curves measured in 1 T on cooling and heating, the maximal intrinsic value of the magnetic entropy change ($\Delta S_m$) for $\Delta B = 0$-5 T, and the discontinuous changes of the lattice parameters $|\Delta a/a|$ and $|\Delta c/c|$ at the transition for the $Mn_xFe_{1.95-x}P_{1-y}Si_y$ compounds. Note: the thermal hysteresis is not corrected for thermal lag of the equipment that results under the same conditions for Gd in a thermal hysteresis of 0.5 K.

| Composition | | $T_C$ (K) | $T_O$ (K) | $|\Delta S_m|$ (Jkg$^{-1}$K$^{-1}$) | $\Delta T_{hys}$ (K) | $|\Delta a/a|$ (%) | $|\Delta c/c|$ (%) |
|---|---|---|---|---|---|---|---|
| $y = 0.50$ | $x = 1.20$ | 304 | ~ 400 | ~ 31 | 6 | 1.1 | 2.2 |
| | $x = 1.25$ | 285 | ~ 390 | ~ 27 | 2 | 0.8 | 1.7 |
| | $x = 1.30$ | 269 | ~ 380 | ~ 21 | 1 | 0.7 | 1.5 |
| $x = 1.25$ | $y = 0.45$ | 246 | ~ 360 | ~ 33 | 7 | 1.3 | 2.5 |
| | $y = 0.50$ | 285 | ~ 390 | ~ 27 | 2 | 0.8 | 1.7 |
| | $y = 0.55$ | 323 | ~ 420 | ~ 19 | 1 | 0.5 | 1.2 |

Furthermore we observe above $T_C$, non-linear temperature variations of the lattice parameters below a certain temperature $T_O$ (see Fig. 2) while the volume increases linearly with increasing temperature. The obtained values for $T_O$ are listed in Table I. The variations of the lattice parameters imply that the increase in hybridization still continues above $T_C$. Thus, for the Mn-rich compounds the Fe/Mn(3*f*) moments do not actually disappear at $T_C$, but their size significantly drops to a lower value. With further increasing temperature, the Fe/Mn(3*f*) moments gradually decrease and vanish at $T_O$ while the Mn(3*g*) moments prevail above $T_O$. The thermal evolution of these moments leads to the non-linear variations of the lattice parameters via the magneto-elastic coupling. Thus, $T_O$ represents the onset of the moment formation on the 3*f* site.

The following physical picture evolves from our results: Increasing Mn or Si contents leads to a reduction in valence electron concentration because the number of valence electrons of Mn or Si is smaller than that of the substituted element Fe or P, respectively. This will affect the bonding properties of the 3*f* sites. At high temperatures above $T_O$, Mn on the 3*g* sites carries magnetic moments and the 3*f* sites are predominantly in the bonding state. The weakening of the bonding by the introduction of Mn and Si results only in a small expansion along *a* and contraction along *c* direction (see Fig. 2). At lower temperatures, due to ferromagnetic exchange interaction between the Mn(3*g*) moments the Fe/Mn(3*f*) moments experience a local field that supports the non-bonding high-moment state and when the exchange field exceeds a certain value the discontinuous step in lattice parameters occurs. An external magnetic field can also support the non-bonding high-moment state and make the transition happen at higher temperatures (see Fig. 4 for a representative sample). The value of d$T_C$/d$B$ is estimated to be about 3-4 K/T. Here, $T_C$ is determined from extremes of the first derivative of the isofield magnetization with respect to temperature.

## IV. MAGNETO-ELASTIC COUPLING, HYSTERESIS AND MAGNETIC ENTROPY CHANGE

Though the lattice parameters change drastically at the transition, only a small volume increase of about 0.1 % with increasing temperature was observed. That is because the *a* and *c* parameters change in opposite sense. Thus, $\Delta a$ and $\Delta c$ have opposite signs. Note that the magnitude of $\Delta a$ and $\Delta c$ decreases with increasing the Mn and Si content (see Fig. 2). For the Mn$_x$Fe$_{1.95-x}$P$_{0.50}$Si$_{0.50}$ compounds, the $|\Delta a/a|$ and $|\Delta c/c|$ change from 1.1 % and 2.2 % to 0.7 % and 1.5 %, respectively, when *x* increases from 1.20 to 1.30. For Mn$_{1.25}$Fe$_{0.70}$P$_{1-y}$Si$_y$ compounds, the $|\Delta a/a|$ and $|\Delta c/c|$ change from 1.3 % and 2.5 % to 0.5 % and 1.2 %, respectively, with increasing *y* from 0.45 to 0.55 (see Table I).

Concerning the hysteresis, it is found that thermal hysteresis decreases with increasing *x* and *y*. Figure 3 shows that the thermal hysteresis reduces from 6 K down to 1 K with increasing *x* from 1.20 to 1.30, and from 7 K down to 1 K when *y* increases from 0.45 to 0.55. Comparing the data listed in Table I we find that the magnitude of the hysteresis can be correlated with that of the $\Delta a$ and $\Delta c$ steps. The larger the $|\Delta a/a|$ and $|\Delta c/c|$, the larger the hysteresis. It is well-known that the hysteresis originates from the energy barrier of the FOMT. Thus, the energy barrier is high when the lattice parameter difference between the ferro- and paramagnetic phases at the transition is large.

With large magneto-elastic coupling, we expect to obtain a large MCE at the FOMT. In this study, magnetic entropy change ($\Delta S_m$) is derived from magnetization isotherms using the Maxwell relations.[1] Large $|\Delta S_m|$ can be achieved if the field-induced transition is complete, i.e., the whole sample is converted from one state to another. The $\Delta S_m$ observed will not reflect the intrinsic value of the $\Delta S_m$ of the compounds if the field-induced transition is incomplete. Thus, intrinsic $|\Delta S_m|$ obtained from complete field-induced transition is an important

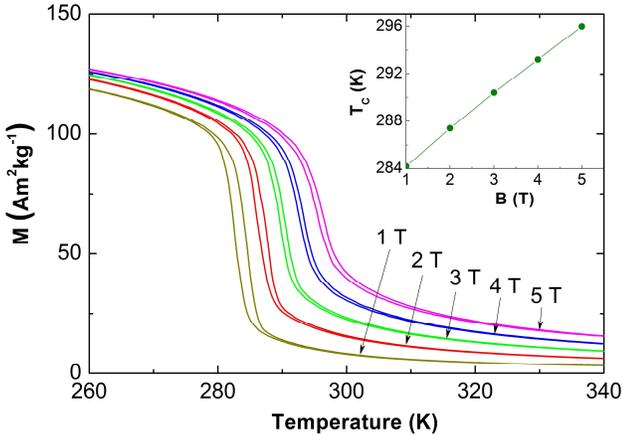

FIG. 4 (Color online). Temperature dependence of the magnetization measured in different magnetic fields (from 1 to 5 T) on cooling and heating for the Mn$_{1.25}$Fe$_{0.70}$P$_{0.50}$Si$_{0.50}$ compound (*x* = 1.25 and *y* = 0.50). The inset shows the field dependence of the transition temperature derived from the *M-T* curves on heating.

parameter to evaluate and compare the size of the MCE of the compounds.

Figure 5a shows the magnetization isotherms measured in increasing field from the paramagnetic state for the sample with $x = 1.25$ and $y = 0.50$ as a representative. Each *M-B* curve was measured after the sample was zero-field cooled from the paramagnetic state at high temperature in order to avoid the coexistence of the ferro- and paramagnetic phases from the previous *M-B* measurement.[18] One can see that a low magnetic field is not sufficient to complete the field-induced transition. There is a critical magnetic field which drives the field-induced transition to completion.[19] The critical magnetic field tends to shift to higher values with increasing the measuring temperature. The values of $\Delta S_m$ as derived from the magnetization isotherms are shown in Fig. 5b. Even though the field-induced transition is complete, the alignment of the magnetic moments may be influenced by thermal fluctuations. A high magnetic field tends to further align the magnetic moments. Thus, the intrinsic $|\Delta S_m|$ increases slowly with increasing magnetic field above the critical magnetic field. However, this increase in $|\Delta S_m|$ above the critical magnetic field is much less compared with that obtained from the field-induced transition. As can be seen from Fig. 5a the field-induced transition becomes less pronounced with increasing temperature. Thus, the intrinsic $|\Delta S_m|$ will be larger at temperatures close to $T_C$ (see Fig. 5b).

Shown in Fig. 6 is the magnetic field dependence of $|\Delta S_m|$ for the $Mn_xFe_{1.95-x}P_{0.50}Si_{0.50}$ and $Mn_{1.25}Fe_{0.75}P_{1-y}Si_y$ compounds at temperatures near the $T_C$s. Obviously, field-induced transitions are complete with increasing field from 0 to 5 T for all the compounds. The $|\Delta S_m|$ for a field change of 0-5 T near the $T_C$s as seen in Fig. 6 can be considered to be approximately the maximal intrinsic $|\Delta S_m|$ of the compounds under the same magnetic field change. We have clear indications that the maximal intrinsic $|\Delta S_m|$ is larger for the compound with lower $x$ or $y$, pointing to the correlation of the size of the MCE with the $|\Delta a/a|$ and $|\Delta c/c|$ (see Table I).

Very recently, Gschneidner *et al.*[19] proposed that the larger the volume change at the transition, the larger the $|\Delta S_m|$. For the Mn-Fe-P-Si compounds displaying a first-order magnetoelastic transition, there are changes in the lattice parameters at the transition but the volume hardly changes. Moreover, the

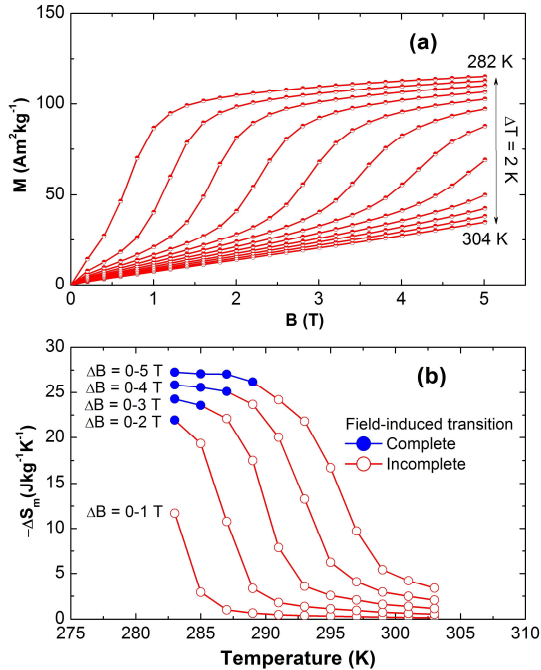

FIG. 5 (Color online). Magnetization isotherms measured in increasing field from the paramagnetic state in the vicinity of the transition temperature (a) and temperature dependence of the magnetic entropy change derived from the magnetization isotherms (b) for the $Mn_{1.25}Fe_{0.70}P_{0.50}Si_{0.50}$ compound ($x = 1.25$ and $y = 0.50$).

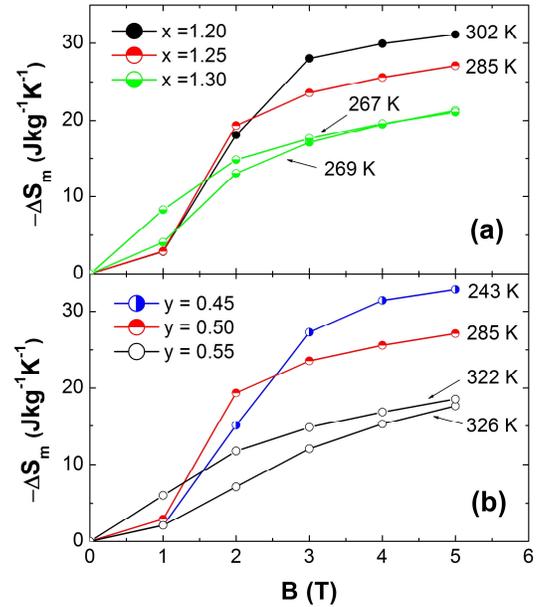

FIG. 6 (Color online). Magnetic entropy change as a function of magnetic field for the $Mn_xFe_{1.95-x}P_{0.50}Si_{0.50}$ (a) and $Mn_{1.25}Fe_{0.70}P_{1-x}Si_x$ (b) compounds near the transition temperatures derived from the magnetization isotherms which were measured in increasing field from the paramagnetic state in the vicinity of the transition temperatures.

changes in the lattice parameters give rise to redistribution of the valence electron and chemical bonding, and simultaneously cause a variation of the magnitude of the magnetic moments, exchange coupling and magnetic order. Thus, the correlation of the $|\Delta S_m|$ with the $|\Delta a/a|$ and $|\Delta c/c|$ is not surprising.

An incomplete field-induced transition gives a weak MCE because there is no phase transition for some part of sample. For magnetic refrigeration applications, the cooling power will be enhanced if complete field-induced transitions are exploited. Since magnetic refrigeration applications often use magnetic fields of less than 2 T produced from permanent magnet, they favor a low critical magnetic field in order to easily drive the field-induced transitions to completion. As mentioned above, the closer the temperature to $T_C$, the smaller the critical magnetic field. Thus, large MCE is always observed in the vicinity of $T_C$. In principle, the critical magnetic field is equal to zero for the field-induced transition at $T_C$ and the maximal $|\Delta S_m|$ is also observed at $T_C$. However, if there is a distribution of $T_C$s, the critical magnetic field must be larger than a certain value in order to be able to drive the field-induced transition to completion (from para- to ferromagnetic phase) for the whole sample with different $T_C$s. Unfortunately, the distribution of the $T_C$s is quite common for FOMT materials because of e.g. inhomogeneous composition.[17,20] Besides, the sudden expansion and contraction of the crystal lattice can create constraints imposed by grain boundaries. Each grain is constrained by its neighbors, leading to the formation of a span of $T_C$s due to inhomogeneity of grain size and shape. The constraints are also obstacles for the FOMT, resulting in an increase in the hysteresis. The distribution of $T_C$s leads to maximal intrinsic $|\Delta S_m|$ at the upper limit of the span of $T_C$s. Within the span of $T_C$s, the field-induced transitions are not complete because they start from the mixed ferro- and paramagnetic states. The $\Delta S_m$ derived from magnetization isotherms using the Maxwell relations should be calculated with great care within the span of $T_C$s.[21]

The composition homogeneity can be improved by changes in preparation technique or heat treatment.[20] Introducing porosity is a possibility for reducing the constraints due to partial removal of the grain boundaries.[22-24] Porous materials should preferably consist of single-crystalline particles with uniform size distribution that will sharpen the transition and enhance the MCE.[23] The porous architecture also improves mechanical properties of the materials and leads to a considerable reduction in hysteresis. The hysteresis reduction has recently been observed by Lyubina et al.[23] in porous La(Fe,Si)$_{13}$ alloys. Sasso et al.[24] also found in Ni$_{55}$Mn$_{20}$Ga$_{25}$ metallic foams that the porous structure reduces the temperature span of the phase transition and increases the MCE.

In the present study, the maximal $|\Delta S_m|$ for the $x = 1.30$ (see Fig. 3a) and $y = 0.55$ (see Fig. 3b) samples with very small hysteresis is estimated to be about 15 Jkg$^{-1}$K$^{-1}$ and 12 Jkg$^{-1}$K$^{-1}$ at 267 K and 322 K, respectively, for a field change of 0-2 T. The magnetic field of 2 T seems to be high enough to drive the field-induced transition to completion (see Fig. 6). The MCE of these samples can still be further enhanced to achieve high cooling power for a magnetic field change of less than 2 T that may be generated by a permanent magnet assembly.[25]

The $|\Delta S_m|$ and hysteresis are both correlated with $|\Delta a/a|$ and $|\Delta c/c|$. Reducing the hysteresis leads to a decrease in the $|\Delta S_m|$. However, small hysteresis can still be obtained while maintaining giant MCE (see Table I). Note that the intrinsic value of the thermal hysteresis is even smaller than those listed in Table I because the thermal lag of the measuring instrument and the influence of the constraints have not been taken into account.

## V. CONCLUSIONS

In summary, a large magneto-elastic coupling caused by competition between moment formation and chemical bonding has been proposed to be at the basis of the FOMT in the hexagonal Fe$_2$P-based materials. By taking advantage of the thermal evolution of the lattice parameters, the increase in magnetic moments upon cooling in the ferromagnetic state which is derived from neutron diffraction measurements can be explained based on the mixed strong and weak magnetism. The formation and gradual development of the magnetic moment on the 3$f$ sites give rise to anomalous thermal expansion in the paramagnetic state. We also observe clear correlation of hysteresis and the $|\Delta S_m|$ with the $|\Delta a/a|$ and $|\Delta c/c|$. Small hysteresis can be achieved without losing giant MCE favorable for magnetic refrigeration applications.

## ACKNOWLEDGMENTS

The authors would like to thank Anton J. E. Lefering, Michel P. Steenvoorde and Bert Zwart (Delft University of Technology) for their help in magnetic and structural measurements, and sample preparation. This work is part of the Industrial Partnership Program of the Dutch Foundation for Fundamental Research on Matter (FOM), and co-financed by BASF Future Business.